\documentstyle[12pt,aasms4]{article}

\def\vol#1  {{{#1}{\rm,}\ }}
\newcommand{\etal}{et~al.~}

\begin{document}

\title{Effects of Weak Gravitational Lensing from Large-Scale Structure on the Determination of $q_0$}
\author{Joachim Wambsganss\altaffilmark{1,3},
Renyue Cen\altaffilmark{1,2}, Guohong Xu\altaffilmark{4} and Jeremiah P. Ostriker\altaffilmark{1}}

\altaffiltext{1} {Princeton University Observatory, Princeton University, Princeton, NJ 08544}
\altaffiltext{2} {Department of Astronomy, University of Washington, Seattle, WA 98195}
\altaffiltext{3} {Astrophysikalisches Institut Potsdam, An der Sternwarte 16, 14482 Potsdam, Germany}
\altaffiltext{4} {Board of Studies in Astronomy and Astrophysics,
University of California, Santa Cruz, CA 95064}

\begin{abstract} 
Weak gravitational lensing by large-scale structure
affects the determination of 
the cosmological deceleration parameter $q_0$.
We find
that the lensing induced dispersions on truly standard candles
are $0.04$ and $0.02$ mag at redshift $z=1$ and $z=0.5$, respectively,
in a COBE-normalized cold dark matter universe with 
$\Omega_0=0.40$, $\Lambda_0=0.6$,
$H=65$km/s/Mpc and  $\sigma_8=0.79$.
It is shown that one would observe
$q_0=-0.44^{+0.17}_{-0.05}$ and
$q_0=-0.45^{+0.10}_{-0.03}$
(the errorbars are $2\sigma$ limits) with
standard candles with zero intrinsic dispersion
at redshift $z=1$ and $z=0.5$, respectively,
compared to the truth of $q_0=-0.40$ in this case, i.e.,
a 10\%  error in $q_0$ will be made.
A standard  COBE normalized $\Omega_0=1$ CDM model
would produce three times as much variance and a mixed
(hot and cold) dark matter model would lead to an intermediate result.
One unique signature of this dispersion effect is its non Gaussianity.
Although the lensing induced dispersion at lower redshift is
still significantly smaller
than the currently best observed (total) dispersion of $0.12$ mag 
in a sample of type Ia supernovae, selected with the 
multicolor light curve shape method,
it becomes significant at higher redshift.
We show that there is an optimal redshift, in the range 
$z\sim 0.5-2.0$ depending
on the amplitude of the intrinsic dispersion of the standard candles,
at which $q_0$ can be most accurately determined.
\end{abstract}


\section{Introduction} 

Any interpretation of observations at cosmological scales
is highly dependent on the density of the universe along
the line of sight to the observed object.
It is precisely for this reason that many types of observations
are made, to {\it determine}
the mean density along the line of sight.
But most of the classical cosmological tests
(Weinberg 1972; Peebles 1980)
have been designed on the assumption that the global
mean density $\bar\rho$ will be measured.
At very high redshift the COBE
observations tell us that the universe was very uniform.
However, we now know that the fluctuations
about this mean are large 
[$(\delta\rho/\bar\rho)_{rms} \ge 1$, for
smoothing scales less than $5h^{-1}$Mpc],
and that even the surface mass densities along different lines of sight
can show large fluctuations
 [$(\delta\Sigma/\bar\Sigma)_{rms} \ge 1$] at low redshift.
 Also gravitational instability theory tells u uss that
the growth of clumpiness in the low to intermediate
redshift range depends sensitively on some 
still largely uncertain cosmological parameters 
such as $\Omega_0$ and $q_0$.
Such a non-uniform matter distribution between our
local observing point and     distant light sources
will affect the properties of the sources 
in two different ways:
the angular size of extended objects is changed (Gunn 1967a) 
and the apparent brightness of a source is affected (Gunn 1967b). 

The rare very large fluctuations when 
$\Sigma\approx \rho_{crit} c H_0^{-1}\approx 10^2 \bar\Sigma$
will produce gravitational lensing of multiple images,
where the source and the observer are connected by more than one null geodesic,
and two or more images of a background quasar 
(first discovered by Walsh, Carswell and Weyman 1979) or galaxy  
(for an impressive recent example see Colley, Turner \& Tyson 1996)
can be seen.
This happens along a trivial fraction of lines of sight estimated
to be a fraction $\sim 3\times 10^{-3}$ from observations
of double quasars.
Precisely the rarity of such events make them sensitive
tests of cosmological models 
(\cite{cgot94}; Kochanek 1995; \cite{wcot95}).
The much more common effect caused by gravitational lensing,
due to much smaller fluctuations in $\Sigma$ (``weak" gravitational lensing),
appears as either
some shape distortion of background galaxies by large
scale matter distribution, primarily in the outskirts of galaxy clusters
(\cite{twv90}; \cite{m91}; \cite{ks93}; for a recent
observational example see \cite{skbfwnb96})
or some change in the apparent brightness of background sources.
In this {\it letter} we study the latter effect:
apparent brightness changes of moderate redshift light sources
caused by the large-scale structure,
showing that weak gravitational
lensing can alter the determination of
the cosmological deceleration parameter $q_0$ in significant ways.
In the context of ad hoc assumptions for the matter
distribution based plausibly on observations
see e.g. Ostriker \& Vietri (1986).
Here we compute such a gravitational lensing 
effect in a concrete manner using specific models for 
the growth of structure.
We apply it to type Ia supernovae as 
``standard candles" used for the determination
of the cosmological deceleration parameter $q_0$.
A complementary analytical study of this effect,
based on some approximations of the cosmic matter distributions
by simple models with adjustable parameters 
was recently done by Kantowski et al. (1995), 

\section{Gravitational Lensing Magnification Distribution}

The details of our ray tracing method to study
the gravitational lensing effects in both strong and weak regimes
can be found in Wambsganss \etal (1996),
and the first results 
focusing on the strong lensing effect
in a standard Cold Dark Matter model have been published 
(Wambsganss \etal 1995).
Here we briefly reiterate the basics.
In order to study the lensing effects of a cosmogonic model
due to the large-scale cosmic structure,
large-scale N-body simulations are used to 
produce contiguous matter distributions at all redshifts. 
For the convenience of data handling (without compromising
the wanted accuracy),
we in practice approximate such a contiguous three dimensional
matter distribution
by a stack of about sixty
two dimensional surface matter planes (lens planes)
filling up the redshift range between observer
and sources at high redshift.
This treatment of the matter in the universe in many lens planes
acts effectively as three dimensional gravitational lensing. 
Then we follow a large number of light rays from
an observer at $z=0$ through many lens planes up to a source plane
at high redshift.
In each lens plane the deflection of the light rays
is determined due to all the matter in this plane.

To quantify the effects,
we use a very high resolution
simulation of a cold dark matter model
with $\Omega=0.40$, $\Lambda_0=0.6$,
$H_0=65$km/s/Mpc and $\sigma_8=0.79$ (LCDM)
(normalized to first year COBE; Smoot \etal 1992).
Note that this model produces both
abundances of rich clusters of galaxies at present epoch
and large-scale structure consistent with observations
(Bahcall \& Cen 1992; 
Cen \& Ostriker 1994;
Peacock \& Dodds 1994;
Cen 1996; Eke, Cole \& Frenk 1996).
The simulation is run using
the TPM (\cite{x95}) N-body code
with a box size of $80h^{-1}$Mpc.
A PM grid of $512^3$ points and subcell 
resolution extension (Tree resolution)
of $30$ are used to give a total
nominal dynamic range of $15370$, translating to the
nominal resolution of $5h^{-1}$kpc or a true
resolution of $\sim 13h^{-1}$kpc.
Since what we are interested in here is weak gravitational lensing
effect, both the achieved resolution on small scales and
the large volume of the simulation is adequate.
[Note that for the present purpose
hydrodynamic simulations (e.g. \cite{c92})
are not needed, for baryonic matter does not
contribute significant amount of mass     until  
one reaches galactic scales where cooling causes
high condensation of the baryons.]
This large-scale simulation
has sufficient dynamic range so that we
do not need a ``convolution" step (Wambsganss \etal 1996). 

We fill the universe with square planes
of $8 h^{-1}$Mpc$\times 8 h^{-1}$Mpc comoving (with a thickness
of comoving size $80 h^{-1}$ Mpc),
produced by projecting rectangular cylinders
of $8 h^{-1}$Mpc$\times 8 h^{-1}$Mpc$\times 80h^{-1}$Mpc,
which in turn are
taken from the $L = 80H^{-1} $Mpc simulation boxes at different redshifts and at random angles.
Each of these lens planes has $800\times 800$ pixels           of size
$10\times 10h^{-2}$kpc$^2$.
We follow an originally regular grid of light rays through the lens
planes up to various source redshifts, considering the deflection
of each ray in each plane by all the matter of the plane.
As a result we obtain the positions of the light rays in 
the source plane in a field of angular size of about $(6 \hbox{arcmin})^2$.
With the simple relations
between these positions and those in the sky plane,  we
obtain properties like shear, effective surface mass density, and --
what is important for the purposes here -- the magnification as a
function of position in the source plane 
(cf. \cite{sef92}; \cite{wco96}).

A total of 80 independent stacks of planes 
are sampled (``lines of sight"), each with 800$^2$
positions, and we determine
the magnification distribution  $f(\mu)d\mu$
for sources at redshifts $z = 0.5$ and $z = 1.0$ (see Figure 1). 
The $z=1$ results are shown as thick solid curves and
the $z=0.5$ results are shown as thin solid curves.
The curves in the top panel describe
the probability that a source is (de-)magnified by $\mu$,
in the lower panel the cumulative probability is displayed:
$f( >     \mu)$. 

Note that the median of the cumulative distribution
is left of $\mu = 1$: most sources will be slightly demagnified.
The average magnification is of course 
$\mu = 1$ (i.e. the same as 
if all matter were smeared out), and a small number of sources 
is moderately magnified, with a very small high magnification tail.
There is a minimum value of the 
magnification for each redshift, this corresponds basically to an ``empty
beam" case,  in which no matter is inside the ray bundle.
The $f(     > \mu)=(97.5\%, 50\%, 2.5\%)$ points are 
$\mu=(0.951,0.983,1.101)$ and 
$\mu=(0.978,0.993,1.034)$, respectively, at
redshifts $z=1$ and $z=0.5$.  
We will use these magnification distributions  
to compute the effects on the determination of 
$q_0$. 
To set the context, let us now discuss
the particular observations concerning
the type Ia Supernovae.

\section{Type Ia Supernovae as Standard Candles: Determination of q$_0$}

Recently, at least a dozen supernovae at redshifts of $z=0.3$
or beyond have been detected (Perlmutter et al. 1995a,b,c;
Garnavich \& Challis 1996a,b), 
most of them of or consistent with type Ia.
They can be used
to determine the 
Hubble constant $H_0$ and 
the cosmological deceleration parameter $q_0$
(Perlmutter et al. 1995a,1996; Riess, Press \& Kirshner 1995).
Currently the most distant supernova detection is at $z = 0.65$
(Perlmutter et al. 1995c).
One assumes knowledge of the absolute brightness of
these supernovae or a subset with additional characteristic features, 
such as
``well-measured light curves" and ``not being unusually red or
spectroscopically peculiar" (\cite{bm93}; \cite{v95};
Perlmutter et al. 1995a), 
known ``minimum ejection velocities of calcium" 
(Fisher et al. 1995),
or ``Light Curve Shapes" (LCS; \cite{p93}; \cite{rpk95}).

For all SN Ia in such a subset
it is assumed that they are standard candles
or their brightness depends in a simple way on an observable
parameter. 
The observed spread, e.g., of 27 ``normal" SN Ia's as
reported by  Perlmutter et al. (1995a) is 
$\sigma_{M_B} = 0.3$ mag in B or 
$\sigma_{M_V} = 0.25$ mag in V, with the intrinsic dispersion expected
to be even smaller. 
Riess, Press, \& Kirshner (1996)
use a multicolor light curve shape method (MLCS)
and are able to select a subset of type Ia supernovae
with a dispersion of only 0.12 mag.
Small intrinsic dispersions 
make SN Ia excellent candidates for measuring
the fundamental cosmological parameters including
Hubble constant $H_0$ and 
the deceleration parameter $q_0$
(for a recent review on the subject see \cite{bt92}).
It is plausibly assumed that there is no systematic 
trend of the intrinsic luminosity of the supernovae
with increasing redshift
(although since the mean metallicity may be 
expected to be redshift dependent,
this assumption should be examined).

Perlmutter et al. (1995a, 1996) 
have shown that it
is possible to determine the value of the cosmological parameter
$q_0$ by detecting supernovae Type Ia at high redshifts 
($z \ge 0.5$). 
We, however, will show that gravitational lensing of the
standard candles by the intervening large-scale matter
distributions causes a systematic, though not large,
underestimate of the
true $q_0$, even if the standard candles are perfect
(i.e., their intrinsic luminosity dispersion is zero).
More importantly, it places     fundamental lower limits 
on the dispersion of the derived $q_0$.

Now, assuming that type Ia supernovae are perfect standard
candles with zero luminosity dispersion,
we can use the magnification distribution shown in Figure 1
to compute the effect of gravitational lensing on
making them appear to be imperfect standard  candles.
In Figure 2, the thin dotted curve indicates
the relationship between redshift (y-axis)
and distance modular, $m-M$ (x-axis)
[see equation (13.30) of Peebles 1993].
Only a small redshift range near z=1.0 and z=0.5, respectively,
in top and bottom panels is shown to best read the effects.
If the matter in the universe were truly smooth,
perfect standard candles would lie on the thin dotted curve.
In both panels,
the three solid dots represent 97.5\%, 50\%, and 2.5\% of
luminosity distribution (converted to distance
modular) of the type Ia supernovae,
when the gravitational lensing effect by
large-scale matter distribution is included.
To quantify the effect in terms of introducing uncertainties
in determining $q_0$, we show in Figure 2 three additional curves:
left thick solid curve, thin solid curve
and right thick solid curve
correspond to three different
universes with $q_0=(-0.27, -0.44, -0.49)$
and $q_0=(-0.35, -0.45, -0.48)$, respectively,
in the top and bottom panels
(a flat cosmological model with a non-zero 
cosmological constant is assumed for simplicity and consistency).
These three curves represent the best fits to the solid dots.
We see that even perfect standard candles with zero dispersion
in a   $$q_0=-0.40$$ universe
will be observed/interpreted as if they reflected a 
\begin{equation}
q_0=-0.44^{+0.17}_{-0.05} 
\end{equation}
\noindent universe for $z_s=1.0$, and a
\begin{equation}
q_0=-0.45^{+0.10}_{-0.03} 
\end{equation}
\noindent universe for $z_s=0.5$,
due to only the gravitational lensing effect by large-scale structure.
Here the value of $q_0$ is the median one, and 
      the errorbars represent $2\sigma$ limits.
Note that the lensing induced
dispersion on $q_0$ is non-Gaussian, which
was pointed out in an earlier study on the same subject
by Dyer \& Oattes (1988).
A similar calculation by Kantowski \etal (1995)
gave a larger, 50\% effect on the median value of
$q_0$, compared to a 10\% effect found here.
The difference is likely due to different models adopted here and
in their study,
in that we use an observationally favored, realistic cosmological
model while their model is highly simplified and not observationally
constrained.

\section{Discussion}

We use a spatially flat cold dark matter model
with $\Omega_0=0.40$, $\Lambda_0=0.6$, 
$H_0=65$km/s/Mpc and $\sigma_8=0.79$,
to compute the effect of gravitational lensing
by large-scale matter distribution on the determination of $q_0$.
This model
agrees with observations on the abundances of rich clusters of galaxies
and large-scale structure as well as is normalized to COBE.
It is shown that, even with perfect standard candles with zero
dispersion, the observed dispersion, after being convolved
with the lensing effect, is sizable with
a dispersion of $0.04$ and $0.02$ mag
at redshifts $1$ and $0.5$, respectively, in the LCDM universe.
Moreover, since the distribution is asymmetric,
the median observed value of $q_0$ is lowered
with resulting values of $q_0=-0.44^{+0.17}_{-0.05}$ 
and $q_0=-0.45^{+0.10}_{-0.03}$
for $z_s=1.0$ and $z_s=0.5$, respectively
(the errorbars are $2\sigma$ limits),
compared to the truth of $q_0=-0.40$ in this case, i.e.,
a 10\% error in $q_0$ will be made.
The non-Gaussian signature clearly distinguishes this lensing
dispersion from other possible broadening of the 
supernova Ia luminosity distribution.
It is worth noting that, although
these dispersions are still smaller 
than the current observed total dispersions of 
some carefully selected subsamples of type Ia supernovae 
at lower redshift 
[for example, $0.12$ mag from Riess \etal (1996)
for SN Ia at redshift between $0.007 < z < 0.1$,
$0.21$ mag from \cite{p96} 
for SN Ia at redshift between $0.35 < z < 0.5$],
it will become increasingly important
for supernovae (or any other standard candles with comparable
or small intrinsic luminosity dispersions) at higher redshift.
Similarly, the lensing induced dispersion will become more
important if the observational errors can be further reduced.

The size of this dispersion, especially the upward 
dispersion of $q_0$,
will depend on the degree of clumpiness 
of the large-scale matter distribution as well as 
on the values of $\Omega_0$ and $q_0$.
We also examined the standard $\Omega_0=1$ cold dark matter (SCDM)
normalized to the first year COBE ($\sigma_8=1.05$),
and found that the lensing induced dispersions are
three times as large as those in the LCDM model presented here.
This  is not surprising 
since it is known that the SCDM model overproduces
the present day rich cluster numbers by a large factor 
(\cite{bc92}; \cite{wef93}), while the adopted LCDM model appears 
to agree
with most observations (\cite{bc92}; 
Ostriker \& Steinhardt 1995;
\cite{c96}; Eke \etal 1996).
We estimate from other calculations
that a hot and cold dark matter model (HCDM) with
$\Omega_0=1$
would produce an intermediate result,
perhaps twice 
as large as the $\sigma=0.04$ quoted here for the
LCDM model.

Even though the differences in the geometry of
the universe for cosmological models with different $q_0$
are increasingly larger at higher redshifts 
(precisely for this reason it is generally thought that
higher redshift sources are better suited to determine $q_0$),
the distortions/dispersions caused by gravitational lensing 
on $q_0$ also become larger at higher redshifts.
Furthermore, gravitational lensing due to  
matter fluctuations on scales smaller than what is
considered here, such as 
lensing by individual galaxies and
microlensing by stars or brown dwarfs
(Linder, Schneider, \& Wagoner 1988;
Rauch 1991), will further increase
the amplitudes of the dispersion.
Therefore, it is not clear whether one gains or loses
by resorting to very high redshift sources, or what the 
``optimal" redshift range for such studies is.

We show, in Figure 3, 
$\eta (z) \equiv {d q_0\over d (m-M)} \sqrt{\Delta M_{instrinsic}^2 + 
\Delta m_{lensing}^2(z)}$
as a function of redshift $z$,
where $\Delta M_{intrinsic}$ denotes the intrinsic magnitude
dispersion of the standard candles,
and $\Delta m_{lensing}(z)$ represents the lensing induced
dispersion in the apparent magnitude
(we have assumed that lensing induced magnitude
dispersions due to other, smaller scale structures will increase 
those due to large-scale structure alone by a factor of two).
$\eta (z)$ quantifies the dispersion in the determined
$q_0$ given the observed dispersion in the apparent magnitude.
Four cases for $\Delta M_{intrinsic}=(0.20, 0.12, 0.06, 0.03)$
mag are plotted (again, a flat cosmological model with a non-zero 
cosmological constant is assumed for simplicity and consistency).
We see that 
one benefits greatly by going from redshift zero to about redshift 0.5;
going further in redshift does not grant a significant gain.
In fact, we find that
the optimal redshift of SNe Ia for determining $q_0$
is $(2.0, 1.4,0.70,0.45)$, respectively, for the four assumed 
$\Delta M_{intrinsic}$ values.
Note that this effects applies to any set of standard candles;
it is not limited to Type Ia supernovae.

Finally, we point out that, if there exists
a set of truly standard candles 
with a negligibly small intrinsic luminosity dispersion,
it should be possible to use them to
directly probe the gravitational lensing effect 
discussed here.
But, on the other hand,
we warn that even quite tiny systematic changes
in either the intrinsic luminosity of the
sources with increasing look-back time
or a redshift dependent variance could easily
give quite incorrect results for both $q_0$ and
the detected lensing.

\acknowledgments
The work is supported in part
by grants NAG5-2759, AST91-08103 and ASC93-18185.
Discussions with P. H\"oflich, C. Hogan, 
B. Kirshner, G. Lake, S. Perlmutter and C. Stubbs 
are gratefully acknowledged.
RC wants to thank G. Lake and University of Washington for 
the warm hospitality, and financial support 
from the NASA HPCC/ESS Program during a visit when 
much of this work was done.

\newpage
\section*{Figure Captions}

\figcaption[FLENAME]{
         Magnification distribution due to gravitational lensing  by
large-scale matter distributions in a $\Lambda$-dominated, 
COBE (first-year) normalized  flat cold dark 
matter model 
for sources at redshifts $z = 0.5$ (thick lines) and $z = 1.0$ (thin lines).  
The curves in the top panel describe
the probability $f(\mu)$that a source is (de-)magnified by $\mu$,
in the lower panel the cumulative probability is displayed:
$f(>\mu)d\mu$. 
\label{fig1}}

\figcaption[FLENAME]{
       The thin dotted curve shows the location of perfect standard
       candles in the distance modulus ($m-M$) - redshift plane
for a $q_0=-0.40$ universe.
The three solid dots represent 2.5\%, 50\%, and 97.5\% of
luminosity distribution (converted to distance
modular) of the type Ia supernovae at $z=1$ (top panel)
and $z=0.5$ (bottom panel), 
respectively, after the 
large-scale structure lensing effect is considered.
The left thick solid curve, thin dotted curve
and right thick solid curve
correspond to three different
universes with $q_0's$ as indicated.
(a flat cosmological model with a non-zero 
cosmological constant is assumed for simplicity and consistency).
These three curves represent the best fits to the solid dots.
\label{fig2}}

\figcaption[FLENAME]{
We show $\eta(z)$ (see text), which
is proportional to the dispersion in the determined $q_0$,
as a function of redshift $z$.
Four cases for the intrinsic magnitude dispersion of the standard candles
                $\Delta M_{intrinsic}=(0.20, 0.12, 0.06, 0.03)$ mag are plotted
(a flat cosmological model with a non-zero 
cosmological constant is assumed for simplicity and consistency).
Note that, for a given intrinsic magnitude dispersion 
of a sample of standard candles, the minimum of $\eta(z)$ denotes the optimal redshift 
at which $q_0$ can be most accurately determined,
using any type of standard candles.
\label{fig3}}

\vfill\eject

\clearpage
\begin{thebibliography}{DUM}
\bibitem[Bahcall \& Cen 1992]{bc92} Bahcall, N.A., \& Cen, R. 1992, \aplett, 398, L81 
\bibitem[Branch \& Miller 1993]{bm93} Branch, D., \& Miller, D.L. 1993, \apj, 405, L5
\bibitem[Branch \& Tammann 1992]{bt92} Branch, D., \& Tammann, G.A. 1992, ARAA, 30, 359
\bibitem[Cen 1992]{c92} Cen, R. 1992, \apjs, 78, 341
\bibitem[Cen 1996]{c96} Cen, R. 1996, \apj,  submitted
\bibitem[Cen \& Ostriker 1994]{co94} Cen, R., \& Ostriker, J.P. 1994, \apj, 429, 4
\bibitem[Cen \etal 1994]{cgot94} Cen, R., Gott, J.R., III, Ostriker, J.P., \& Turner, E.L. 1994, \apj, 423, 1
\bibitem[Colley \etal 1995]{ctt95} Colley, W. N. , Turner, E.L, Tyson, J.A. 1995, \apjl,  461, L83 
\bibitem[Dyer \& Oattes 1988]{do88} Dyer, C.C., \& Oattes, L.M. 1988, \apj, 326, 50
\bibitem[Eke, Cole, \& Frenk 1996]{ecf96} Eke, V.R., Cole, S., \& Frenk, C.S. 1996, preprint
\bibitem[Fisher \etal 1995]{fbhk95} Fisher, A., Branch, D., H\"oflich, P., Khokhlov, A. 1995, \apj, 447, L73 
\bibitem[Garnavich \& Challis 1996a]{gc96a} Garnavich, P., \& Challis, P. 1996a,IAU Circular No. 6332 
\bibitem[Garnavich \& Challis 1996b]{gc96b} Garnavich, P., \& Challis, P. 1996b,IAU Circular No. 6358 
\bibitem[Gunn 1967a]{g76a} Gunn, J.E. 1967a, \apj, 147, 61 
\bibitem[Gunn 1967b]{g76b} Gunn, J.E. 1967b, \apj, 150, 737 
\bibitem[Kaiser \& Squires 1993]{ks93} Kaiser, N., \& Squires, G. 1993, \apj, 404, 441 
\bibitem[Kantowski, R., Vaughan, T., \& Branch 1995]{kvb95} Kantowski, R., Vaughan, T., \& Branch, D. 1995 \apj, 447, 35
\bibitem[Kochanek 1995]{k95} Kochanek, C.S. 1995, \apj, 453, 545
\bibitem[Linder, Schneider, \& Wagoner 1988]{lsw88} Linder, E.V., Schneider, P., \& Wagoner, R.V. 1988, \apj, 324, 786
\bibitem[Miralda-Escude 1991]{m91} Miralda-Escude, J. 1991, \apj, 370, 1
\bibitem[Ostriker \& Steinhardt 1995]{os95} Ostriker, J.P., \& Steinhardt, P. 19
95, \nat, 377, 600
\bibitem[Ostriker \& Vietri 1986]{ov86} Ostriker, J.P., \& Vietri, M. 1986, \apj, 300, 68
\bibitem[Peacock \& Dodds 1994]{pd94} Peacock, J.A., \& Dodds, S.J. 1994, \mnras, 267, 1020
\bibitem[Peebles 1980]{p80} Peebles, P.J.E. 1980, The Large-Scale Structure of the Universe (Princeton: Princeton University Press)
\bibitem[Peebles 1993]{p93} Peebles, P.J.E. 1993, Principles of Physical Cosmology (Princeton: Princeton University Press)
\bibitem[Perlmutter 1995a]{p95a} Perlmutter, S. \etal 1995a, \apj, 440, L41 
\bibitem[Perlmutter 1995b]{p95b} Perlmutter, S. \etal 1995b, IAU Circular No. 6263 
\bibitem[Perlmutter 1995c]{p95c} Perlmutter, S. \etal 1995c, IAU Circular No. 6270 
\bibitem[Perlmutter 1996]{p96} Perlmutter, S. \etal 1996, preprint
\bibitem[Phillips 1993]{p93} Phillips, M. 1993, \apj, 413, L105
\bibitem[Rauch 1991]{r91} Rauch, K.P. 1991, \apj, 374, 83
\bibitem[Riess \etal 1995]{rpk95} Riess, A.G., Press, W.H., Kirshner, R.P. 1995, \apj, 438, L17
\bibitem[Riess \etal 1996]{rpk96} Riess, A.G., Press, W.H., Kirshner, R.P. 1996,preprint
\bibitem[Schneider \etal 1992]{sef92} Schneider, P., Ehlers, J., \& Falco, E.E. 1992, {\it Gravitational Lensing} (Springer Verlag, Berlin)
\bibitem[Smoot \etal 1992]{s92} Smoot, G.F., \etal 1992, \aplett, 396, L1
\bibitem[Squires \etal 1996]{skbfwnb96} Squires, G., Kaiser, N., Babul, A., Fahlman, G., Woods, D., Neumann, D.M., \& Boehringer, H. 1996, \apj, 461, 572
\bibitem[Tyson \etal 1990]{twv90} Tyson, J.A., Wenk, R.A., \& Valdes, F. 1990, \apj, 349, L1
\bibitem[Vaughan \etal 1995]{v95} Vaughan, T., Branch, D., Miller, D., \& Perlmutter, S. 1995, \apj, 439, 558
\bibitem[Walsh \etal 1979]{wcw79} Walsh, D., Carswell, R.F., Weymann, R.J. 1979, \nat, 279, 381
\bibitem[Weinberg 1972]{w72} Weinberg, S. 1972, Gravitation and Cosmology (New York: Wiley)
\bibitem[Wambsganss \etal 1995]{wcot95} Wambsganss, J., Cen, R., Ostriker, J. P., \& Turner, E. L.  1995, Science, 268, 274 
\bibitem[Wambsganss \etal 1996]{wco96} Wambsganss, J., Cen, R., \& Ostriker, J. P. 1996, preprint
\bibitem[White, Efstathiou, \& Frenk 1993]{wef93} White, S.D.M, Efstathiou, \& Frenk, C.S. 1993, \mnras, 261, 1023
\bibitem[Xu 1995]{x95} Xu, G. 1995, \apjs, 98, 355
\end{thebibliography}
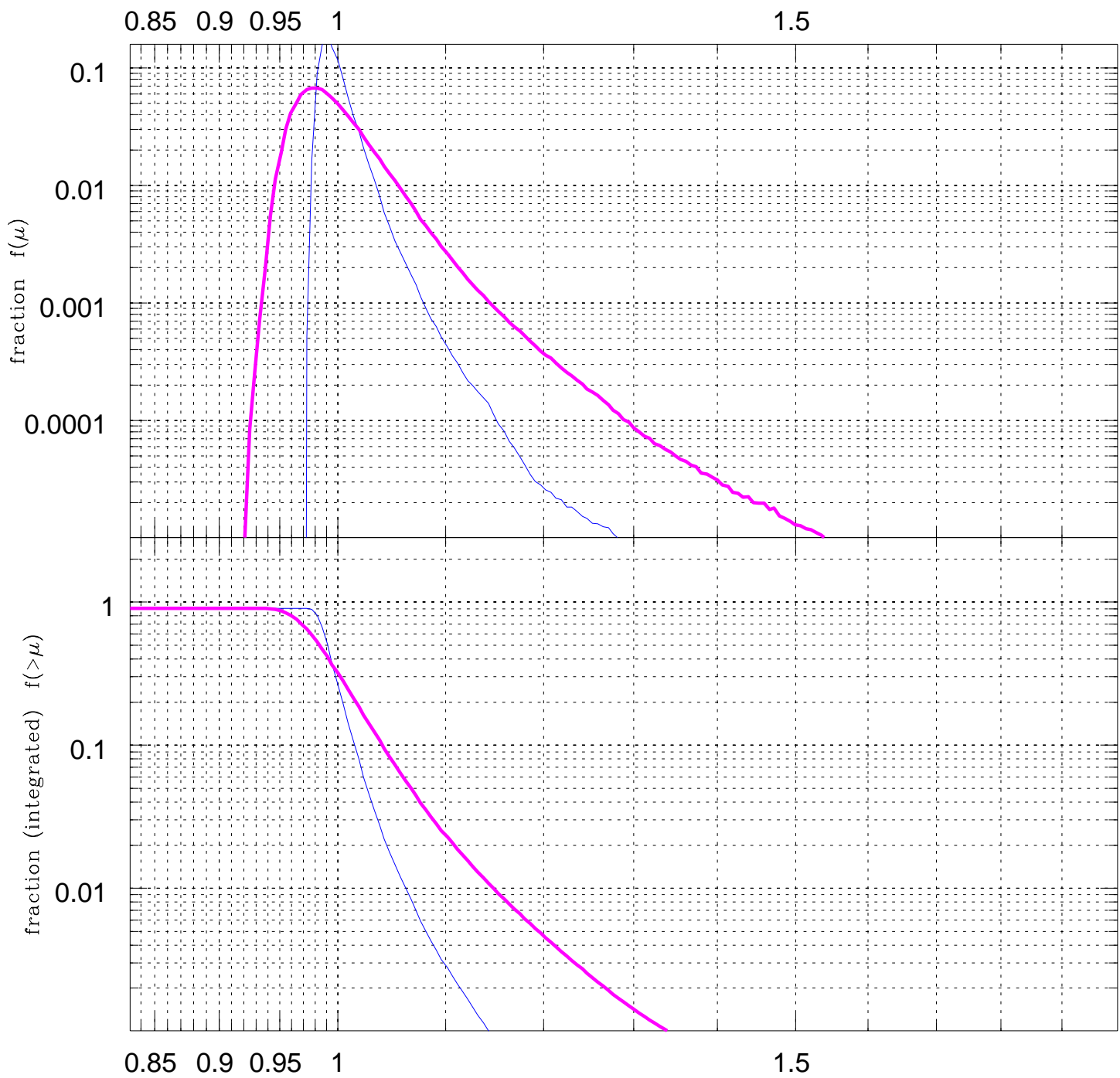
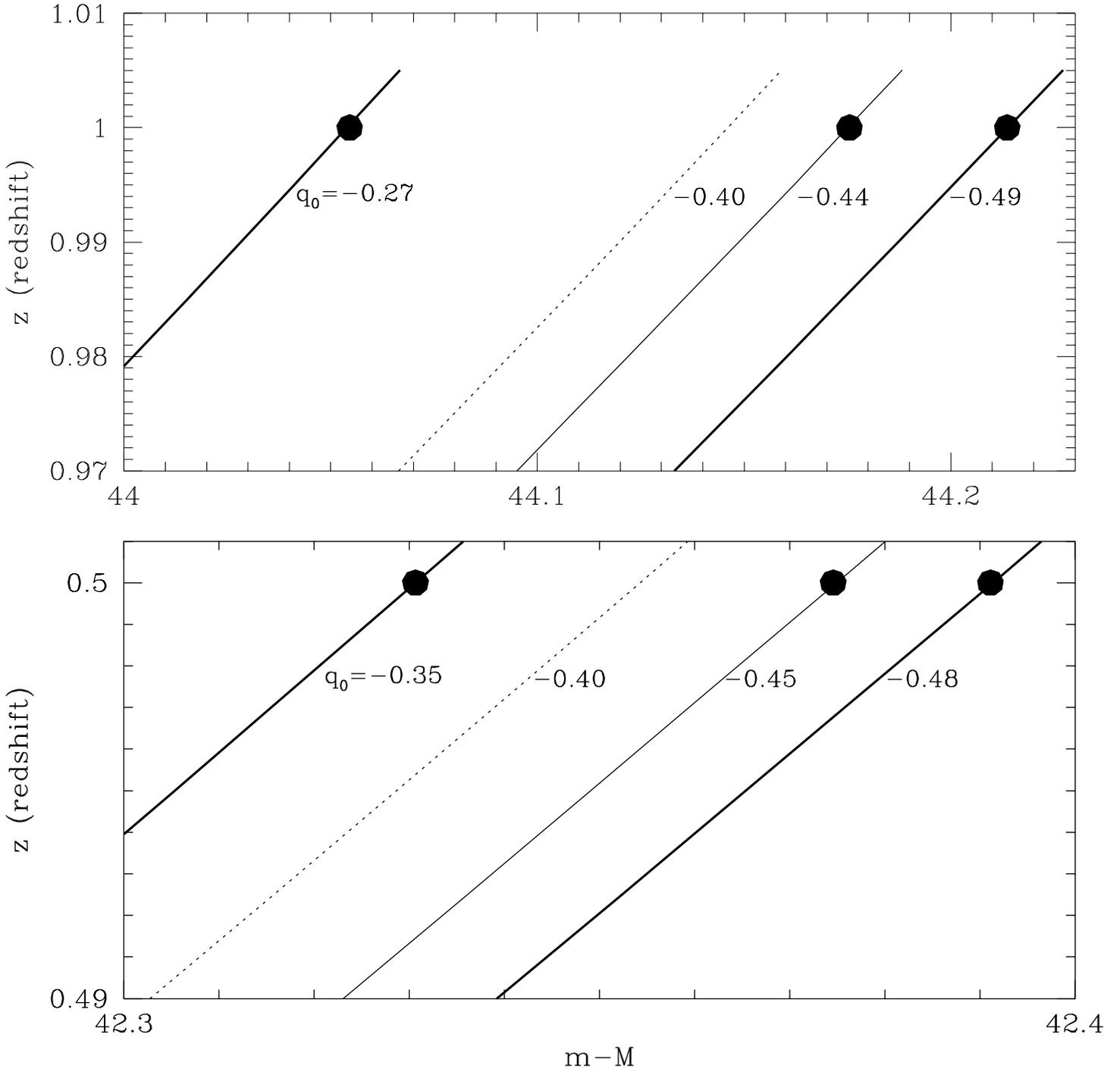
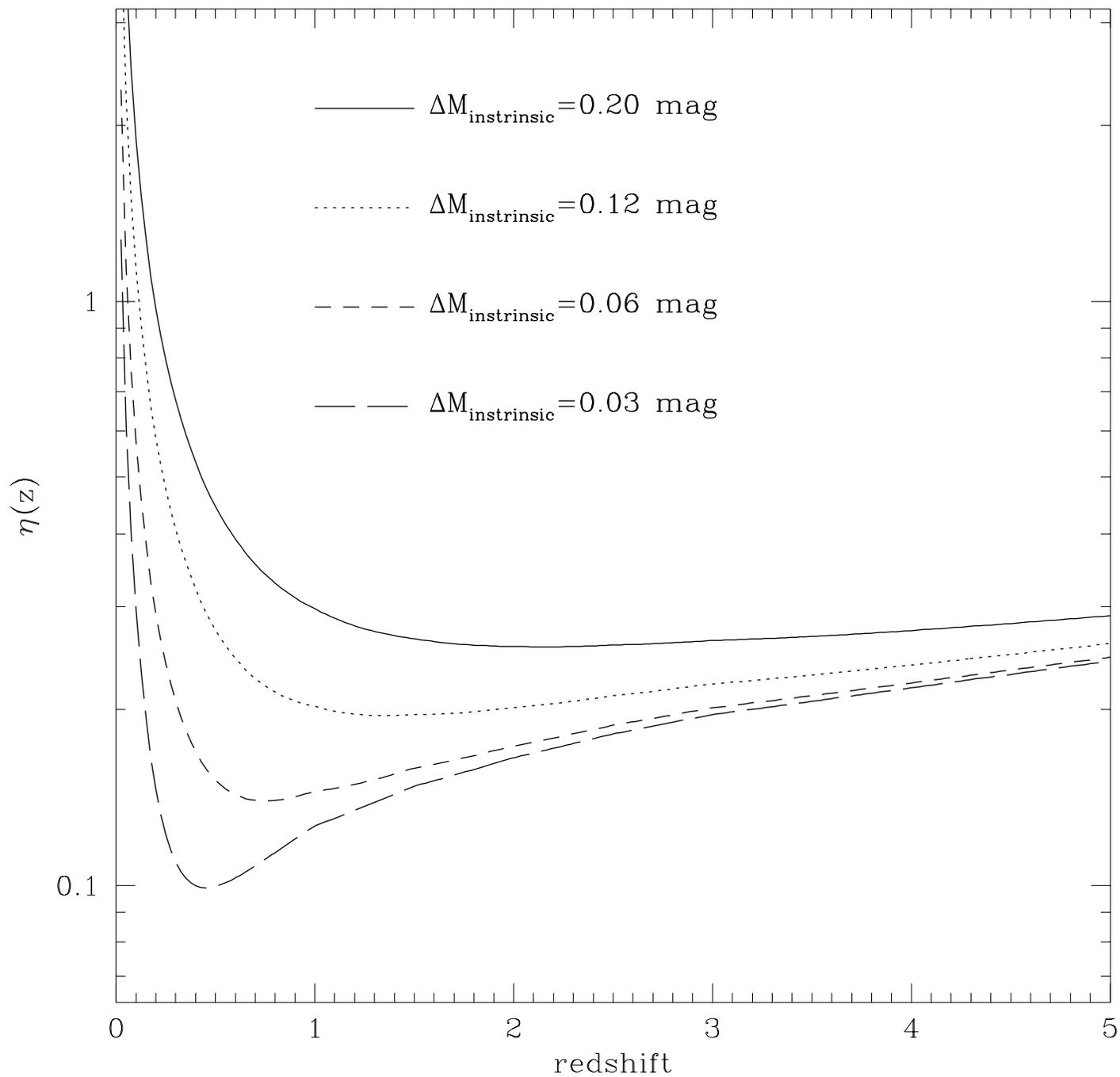
\end{document}